# Formation of Frustrated Charge Density Waves in Kagome Metal LuNb$_6$Sn$_6$

F. Z. Yang[1,*], X. Huang[2,*], Hengxin Tan[3,*], A. Kundu[4], S. Kim[4], M. Thinel[2,5], J. Ingham[2], A. Rajapitamahuni[4], C. Nelson[4], Y. Q. Cai[4], E. Vescovo[4], W. R. Meier[6], D. Mandrus[7], Brenden R. Ortiz[1], A. N. Pasupathy[2,8†], Binghai Yan[3‡], H. Miao[1§]

[1]*Materials Science and Technology Division, Oak Ridge National Laboratory, Oak Ridge, Tennessee 37831, USA*

[2]*Department of Physics, Columbia University, New York, NY, 10027, USA*

[3]*Department of Condensed Matter Physics, Weizmann Institute of Science, Rehovot 7610001, Israel*

[4]*National Synchrotron Light Source II, Brookhaven National Laboratory, Upton, New York 11973, USA*

[5]*Department of Chemistry, Columbia University, New York, NY, 10027, USA*

[6]*Department of Materials Science and Engineering, The University of Tennessee, Knoxville, Tennessee 37996, USA*

[7]*Department of Physics and Astronomy, The University of Tennessee, Knoxville, Tennessee 37996, USA*

[8]*Condensed Matter Physics and Materials Science Division, Brookhaven National Laboratory; Upton, NY, USA*

**The charge density wave (CDW), a translational symmetry breaking electronic liquid, plays a pivotal role in correlated quantum materials, such as high-$T_c$ superconductors[1-3] and topological semimetals[4-8]. Recently, CDWs that possibly intertwine with superconductivity and magnetism are observed in various kagome metals[9-13]. However, the nature of CDWs and the role of the Fermi surface (FS) topology in these materials remain an unresolved challenge. In this letter, we reveal the formation of CDWs in the newly discovered kagome metal LuNb$_6$Sn$_6$[13]. We observe a precursor CDW correlation that features a 'yield sign'-like hollow triangle diffuse scattering pattern and nearly complete softening of a flat optical phonon band near $Q_H$=(1/3, 1/3, 1/2). The scattering intensity of the precursor CDW displays divergent behavior as decreasing temperature until $T_{CDW}$=70 K, where a competing CDW at $Q_{CDW}$=(1/3, 1/3, 1/3) emerges. Using scanning tunneling microscopy/spectroscopy, we image the frustrated CDW patterns that show a short phase coherence length $\xi$~ 20 nm in real space. Combined with first principles calculations, our observations support frustrated lattice**

**interactions that are modulated by the FS topology. These results shed light on the interplay between FS and CDW in quantum materials with nearly degenerate structural deformation patterns.**

The intertwined interplay between spin, charge, orbital, and lattice degrees of freedom is a central focus of correlated quantum materials[3]. In a 1-dimensional electronic liquid, the perfect nested Fermi surface (FS) collaborates with electron-phonon coupling giving rise to a charge density wave (CDW) and the associated periodic lattice deformations[14,15]. In 2D and 3D materials, however, the origin of CDWs is under debate between weak coupling mechanisms that are derived from electronic states near the Fermi level[16-19], and strong coupling scenarios based on high-energy electron-electron or electron-phonon interactions, where FS topology are believed to play a minor role[1,20]. This debate spans from strongly correlated electronic systems, such as cuprate high-$T_c$ superconductors[21-23], to weakly interacting materials, including transition metal chalcogenides[24-28] and kagome metals[29-45].

Figure 1**a** and **b** depicts the crystal structure and Brillouin zone of the newly discovered kagome metal $LuNb_6Sn_6$[13]. $LuNb_6Sn_6$ adopts the $HfFe_6Ge_6$-prototype belonging to the broader $AM_6X_6$ class of materials, with space group *P6/mmm* (No. 191) at room temperature. $LuNb_6Sn_6$ behaves very similarly to the previously discovered $ScV_6Sn_6$, with both compounds exhibiting similar density wave order that breaks translational symmetry[12,13]. To remain consistent with the literature on $ScV_6Sn_6$, we will refer to the low-temperature phase using the "CDW" nomenclature. Like $ScV_6Sn_6$[37-39], the density functional theory (DFT) calculated phonon dispersion (Fig. 1**c**) show unstable flat optical phonon mode with an energy minimum at the $\boldsymbol{Q}_H$=(1/3, 1/3, 1/2) point, corresponding Sn1-Lu-Sn1 *c*-axis distortions and leading to a $\sqrt{3} \times \sqrt{3} \times 2$ supercell[35-42]. As shown in Fig. 1**d**, above the CDW transition temperature, $T>T_{CDW}$=70 K, the elastic x-ray scattering reveals two asymmetric precursor CDW superlattice peaks that are shifted away from $\boldsymbol{Q}_H$ along the HH-direction. Interestingly, the true thermodynamic phase transition is associated with $\sqrt{3} \times \sqrt{3} \times 3$ CDW with $\boldsymbol{Q}_{CDW}$=(1/3, 1/3, 1/3) (Fig. 1**e** and Supplementary Materials), indicating unconventional CDW formation mechanism. In this letter, we combine real and reciprocal space probes to uncover the evolution of CDWs in $LuNb_6Sn_6$. We show that the FS topology of $LuNb_6Sn_6$ induces frustrated charge correlations near $\boldsymbol{Q}_H$ that avoid the thermodynamic order at $\boldsymbol{Q}_H$. This pivotal role of the FS arises from the complete softening of flat-

phonon mode in broad momentum space, where the degeneracy of competing CDW patterns is lifted by itinerant electron-mediated effective lattice interactions.

**The electronic structure of LuNb$_6$Sn$_6$**

We first establish the electronic structure of LuNb$_6$Sn$_6$ using angle-resolved photoemission spectroscopy (ARPES). Figure 2 shows the ARPES intensity on LuSn$_2$- and Nb$_3$Sn-terminated surface as revealed by the Sn 4$d$-electron core levels (Figs. 2**b** and 2**g**)[40,42,46-48]. Their corresponding FS topology and band dispersions along high-symmetry lines (labeled by cyan dotted lines in Figs.2**a** and 2**f**) are shown in Figs. 2**a**-2**e** and Figs. 2**f**-2**j**, respectively. Red and blue curves are DFT calculated bulk electronic structures at $k_z = 0$ and $\pi$, respectively. In agreement with DFT calculations, two triangular FSs at the $\bar{K}$ point are observed on both surface terminations, as shown in Figs. 2**a** and 2**f**. On the LuSn$_2$ surface, we find two additional FSs, highlighted by red arrows in Fig. 2**a**, that are only visible in the second Brillouin zone (BZ). Their band dispersions along the $\bar{K}$-$\bar{M}$-$\bar{K}$-$\bar{\Gamma}$ and $\bar{\Gamma}$-$\bar{M}$ direction are respectively shown in Fig. 2**d** and Fig. 2**e**, and indicated by red arrows. The high-intensity side forms a circular hole-like FS, while the low-intensity side gives rise to a larger hexagonal FS. Similar features have also been observed in ScV$_6$Sn$_6$ and GdV$_6$Sn$_6$, and attributed to surface states[40,48]. The blue arrows in Figs. 2**c-e** and 2**h-j** point to two bulk massive Dirac Fermions at the $\bar{K}$ points with binding energies, $E_B$= 140 and 600 meV. The black arrows denote three van Hove singularities located at $E_B$=30, 70, and 530 meV. We find that the experimental binding energy of Dirac points and van Hove singularities are ~80 meV lower than DFT calculations (see Supplementary Material for details), suggesting a non-negligible charge doping on the surface.

**FS driven precursor CDW correlations**

We now focus on the precursor CDW correlations at $T>T_{CDW}$. Fig. 3**a** shows the precursor CDW intensity in the ($H, K, L$=6.5) plane at $T$=85 K>$T_{CDW}$. Interestingly, we observe a 'yield sign'-like hollow triangular pattern centered at the H/H' point in the HK-plane, where one corner and an edge of the triangle intersect with the HHL-plane, resulting in two asymmetric superlattice peaks in Fig. 1**d**. A detailed analysis of the 'yield sign' reveals local intensity maximum at the three corners and centers of the three edges. This intensity distribution is more obvious in the second derivative plot (Fig. 3**b**) of Fig. 3**a**. To understand the origin of precursor CDW, we calculate the bare charge susceptibility, $\chi(q)$, using the DFT calculated band structure. Remarkably, as shown in Fig. 3c and 3d, $\chi(q)$ displays the same hollow triangles centered at $q=Q_H$. Importantly, not only the size, shape, and location of the hollow triangles in $\chi(q)$ are matching experimental observations, but also the local intensity maximum at the corner and the center of edges agree with x-ray data. These observations strongly support a FS-induced precursor CDW.

Figures 3f tracks the temperature dependent intensity of the precursor CDW, $I_p(T)$, at $Q$=(-0.76, -0.76, 6.5). Figure 3g shows the colormap of a line cut along the (H, H, 6.5) direction as function of temperature. Intriguingly, below $T\sim160$ K, $I_p$(T) increases progressively with decreasing temperature and follows $I_p(T) = C_0 - C_1 \ln T \propto \chi(q)/\chi(0)$[18], as revealed in the inset of Fig. 3f. This characteristic temperature dependent evolution of $I_p(T)$ adds another evidence supporting the FS driven precursor CDW.

In the paradigm of Kohn anomaly[16], one would expect sharp phonon softening, which energy minimum momenta forms a 'yield sign' pattern (see Supplementary Materials). Figures 3e and 3h – 3j show inelastic x-ray scattering (IXS) determined phonon dispersion at $T$ = 300, 150, and 90 K. While we observe a complete softening, within the experimental resolution $\Delta E \sim 1.4$ meV, of a flat optical phonon mode along the (H, H, 6.5) direction, this softening is broad in momentum space. This observation suggests that the strong momentum-dependent electron-phonon coupling also plays a key role in LuNb$_6$Sn$_6$. Indeed, as shown in Fig. 3e, while the flat phonon mode also shows large softening at the $L$-point (Q=0.5, 0.5, 6.5), consistent with DFT calculated peak in $\chi(q)$, this softening is incomplete due to weaker electron-phonon coupling at the $L$-point and hence failed to induce a peak in the elastic x-ray scattering. Assuming the precursor CDW intensity arises from the dynamical lattice excitations, the phonon dynamics will yield filled triangles in the x-ray diffraction (see Supplementary Materials) as observed in ScV$_6$Sn$_6$[37,38]. The observation of 'yield sign' pattern in LuNb$_6$Sn$_6$ supports a FS induced static CDW correlations that is beyond the conventional Kohn anomaly picture.

**Competing CDWs**

As shown in Fig. 3f and 3g, the divergent behavior of the precursor CDW intensity is interrupted by the emergence of $\sqrt{3} \times \sqrt{3} \times 3$ CDW at $T=T_{CDW}$. In comparison, Fig. 3k shows the peak intensity $I_{CDW}(T)$ at $Q_{CDW}$= (-0.33, -0.33, 6.33) that increases sharply at $T_{CDW}$. This competing behavior supports competing lattice distortion patterns and is consistent with the broad softening of low-energy flat phonon mode as revealed by IXS and DFT calculations. The inset of Fig. 3**k** and Fig. 3**l** show $\log I_{CDW}(T)$ at $\boldsymbol{Q}_{CDW}$ and the colormap of a scan along (-0.33, -0.33, *L*) direction. These plots reveal another transition step slightly above $T_{CDW}$, suggesting the formation of real space $\sqrt{3} \times \sqrt{3} \times 3$ domains that is about 0.1% total x-ray scattering volume of the low temperature phase. The presence of small fraction of domains is consistent with the first order CDW transition revealed by thermal transport measurement[13].

**Real space evolution of CDWs**

Finally, we image the real space evolution of CDW on the Nb$_3$Sn surface using scanning tunneling microscopy (STM). Figure 4**a** shows the STM topography measured at $T$=7 K $\ll T_{CDW}$. Interestingly, the CDW displays a short coherence length with an average domain size ~20 nm. The fast Fourier transform (FFT) of STM topography, shown in Fig. 4**b**, reveals narrow structural Bragg peaks and broad CDW superlattice peaks corresponding to $\sqrt{3} \times \sqrt{3}$ superstructure. The small CDW domain size is consistent with the x-ray scattering, which shows a Full-Width-at-Half-Maximum (FWHM) of CDW peak ~0.037 r.l.u. along the (H, H) direction (See Supplementary Materials), yielding an in-plane correlation length $\xi$~24 nm in the bulk. Figure 4**c** presents the atomically resolved CDW image that shows $\sqrt{3} \times \sqrt{3}$ supercell with a triangular trimer in each unit cell. These trimers then form a triangular superlattice where the unit vector of the superlattice is parallel to one of the three edges of the triangular trimer. The CDW phase analysis, shown in Fig. 4**d** (see also Methods), reveals in-plane CDW domains with $\pm 2\pi/3$ phase difference. The triangular trimer shape of the CDW as well as the $\pm 2\pi/3$ phase differentiated domains are different from the tri-hexagonal shape of the CDW observed in ScV$_6$Sn$_6$[40] and could suggest to additional symmetry breaking such as unidirectional order[49-51]. Interestingly, the STM topography shows two-types of triangular trimer orientations, and each orientation displays three domains with $\pm 2\pi/3$ phase difference (see Supplementary Materials). The two-types of triangular trimers might be associated with the K and K' instability in the momentum space that are linked by the mirror symmetry of the 2D hexagonal lattice[49-51].

Upon warming up to 80 K>$T_{CDW}$, the in-plane $\sqrt{3} \times \sqrt{3}$ short-range CDW remains robust on the surface as shown in Fig. 4**e**. Interestingly, when a single CDW domain crosses a unit-cell step, a π-phase shift is observed in Fig. 4f, consistent with precursor CDW superlattice intensity peaked at *L=int.*+0.5 in x-ray diffraction. This observation is in stark contrast with the low-temperature CDW domains that deviates significantly from π-phase shift when crossing a unit-cell step (See Supplementary Materials). Such STM topography unambiguously points to the static precursor CDW above $T_{CDW}$ and is consistent with x-ray scattering.

**Discussion:**

The observations of FS driven precursor CDW correlations above $T_{CDW}$ and short-ranged CDW domains with segregated phase-factor make LuNb$_6$Sn$_6$ an ideal platform for intertwined charge and lattice interactions. As we showed in Figs. 3 and 4, both x-ray and STM at $T>T_{CDW}$ support the $L=0.5$ CDW as the leading instability, consistent with the DFT calculations shown in Fig. 1c. However, the true thermodynamic phase transition is associated with the sub-leading $\sqrt{3} \times \sqrt{3} \times 3$ CDW at $L=0.33$. This contradiction can be understood by considering the FS induced charge fluctuations, $\chi(q)$ at $q_z=\pi$, that frustrate the effective lattice coupling akin to the RKKY interaction in spin systems[52-54]. We emphasize that the low-energy flat phonon mode with strong momentum dependent EPC, as shown in Fig.3, is critical for this analogy. This is because only under this condition, the local Sn-Sn distortions are spatially isolated and nearly free to move in a way similar to the *f*-electron RKKY systems. The FS-mediated effective interaction is thus the dominant factor to determine the CDW pattern. This picture is also consistent with the STM observation of short-ranged CDW domains as the phase coherence length of FS mediated interactions are sensitive to both point and line defect[55,56]. Recently, it has been proposed that the frustrated lattice interactions can be mapped onto a real space $J_1$-$J_2$ spin model[36,41]. Although a fitting of the 'yield sign' pattern using a real-space model may still be possible, as we showed in Fig. 3, the shape, size, and the intensity distributions are naturally derived from the FS topology, providing a microscopic picture for the experimental observations.

It is also interesting to note that while the thermodynamic behaviors of LuNb$_6$Sn$_6$ and ScV$_6$Sn$_6$ display many similarities, the CDWs are different in several key aspects. Firstly, the STM determined topography maps are drastically different between these two materials, suggesting different lattice distortions. This is also evidenced in the elastic x-ray diffractions that show different diffraction patterns at $T> T_{CDW}$. Secondly, the CDW in ScV$_6$Sn$_6$ is six-fold symmetric and long-range ordered at the base temperature without domains across hundred nanometers[40], where

LuNb$_6$Sn$_6$ shows short-ranged domains with different phases. A possible scenario is that the higher density of point defects in LuNb$_6$Sn$_6$ enhances FS effects through *e.g.,* Friedel oscillations. Moreover, the presence of degenerated CDW configurations observed in LuNb$_6$Sn$_6$ is also expected to yield short phase coherence length.

The observation of FS induced CDW frustration may also shed light on puzzles in correlated and topological quantum materials, such as cuprates[22,23] and (Eu, Sr)Al$_4$[8,57]. In these materials, the high-energy electron-electron[21,22] and electron-phonon interactions[8] favors competing and degenerate CDW instabilities, leaving FS topology as the key parameter to determine the CDW patterns.

In summary, combining real and reciprocal space probes with DFT calculations, we determined the evolution of CDWs and uncovered the critical role of FS in flat-phononic kagome metal LuNb$_6$Sn$_6$. Our results help to understand the nature of CDWs in correlated and topological materials and highlight the key role of FS in materials with nearly degenerate CDWs and associated lattice distortion patterns.

## Methods

**Sample preparation and characterizations:**
High-quality single crystals of LuNb$_6$Sn$_6$ were grown by self-flux technique. The growth and characterization are described in Ref. 13.

**ARPES measurements:**
The ARPES experiments were performed at beamline 21-ID-1 of NSLS-II at BNL. The LuNb$_6$Sn$_6$ samples were cleaved *in situ* in a vacuum better than $3\times10^{-11}$ Torr. The measurements were taken with synchrotron light source and a Scienta-Omicron DA30 electron analyzer with a beam size ~ 1 $\mu$m. The total energy resolution of ARPES measurement is approximately 15 meV. The sample stage was maintained at 30 K throughout the experiments.

**X-ray scattering measurements:**
The single crystal elastic X-ray diffraction was performed at the integrated *in situ* and resonant hard X-ray studies (4-ID) beamline of National Synchrotron Light Source II (NSLS-II). The photon energies, which is selected by a cryogenically cooled Si(111) double-crystal monochromator, were 6.977 keV. The sample is mounted in a closed-cycle displex cryostat in a vertical scattering geometry. The incident X-rays were horizontally polarized, and the diffraction was measured using a silicon drift detector. All the data were collected with a polarization analyzer in $\sigma$-$\sigma$ channel.

We also performed resonant elastic x-ray scattering at Lu $L_3$-edge (9.244 keV). As we showed in the supplementary materials, neither the precursor CDW nor the CDW superlattice peak show resonant enhancement at the Lu $L_3$-edge (See Supplementary Materials). This is not surprising as the electronic states near the Fermi level are mainly contributed by Nb $4d$-electrons.

**Inelastic X-ray scattering measurements:**

The inelastic x-ray scattering (IXS) was conducted at beamline 10-ID at the NSLS-II. The highly monochromatic x-ray beam of incident energy E = 9.13 keV ($\lambda$ = 1.36 Å) was focused on the sample with a beam cross-section of ~ 10 × 10 $\mu m^2$. The total energy resolution was $\Delta E$ ~ 1.4 meV (full width at half maximum). The measurements were performed in reflection geometry on sample's (001) surface. Under this geometry, we probed the in-plane CDW superlattice peaks with the $L$-component, where $H$, $K$, and $L$ are indices of the three reciprocal lattice vectors, corresponding to the normal state structure (space group *P6/mmm*) with a=b=5.7461 Å and c=9.5163 Å, respectively. Typical counting times were in the range of 360 to 480 seconds per point in the energy scans at constant momentum transfer $Q$. Detailed descriptions of the fitting procedures employed to extract phonon dispersions from the IXS data are provided in Refs. 8 and 23.

**STM/STS measurements:**

STM data were acquired at temperatures below (7 K) and above (80 K) the transition temperature. Single crystals of LuNb$_6$Sn$_6$ were cleaved in ultra-high vacuum before inserted into the STM scanner. A chemically etched tungsten tip was annealed and calibrated on the Au(111) surface. Spectroscopic data were taken by the standard lock-in technique at 2.451 kHz with bias voltage applied to the sample. Real-space density of states images were acquired by the demodulation of the tunneling current when a small excitation bias voltage was applied to sample with the tip scanning over the sample surface in a constant DC current mode.

**First principles calculations:**

Density functional theory (DFT) calculations were performed using the plane-wave-based Vienna ab initio simulation package (VASP)[58], employing the Perdew-Burke-Ernzerhof (PBE) generalized gradient approximation[59] for electron-electron exchange-correlation interactions. The plane-wave energy cutoff was set to 270 eV, and Lu f electrons were treated as core states in all calculations. The crystal structure was fully relaxed until the residual force on each atom was below 1 meV/Å. A $\Gamma$-centered $k$-point mesh of 10×10×5 was used for structural relaxation, and a denser $k$-mesh of 14×14×8 was applied for self-consistent calculations. The phonon spectrum is computed with the finite-displacement method as implemented in the Phonopy software[60], where a supercell size of $\sqrt{3} \times \sqrt{3} \times 4$ together with a $k$-mesh of 6×6×3 was employed. Spin-orbit coupling (SOC) was included in all calculations except during structural relaxation.

To calculate the Fermi surface and the quasiparticle interference (QPI) pattern, we first constructed a Wannier Hamiltonian using the Wannier90 software[61] interfaced with VASP. The Wannierization process considered Nb $d$ and Sn $p$ orbitals. The Fermi surface was extracted from the 3D discretized band structure calculated on a fine $k$-mesh of 160×160×80. The QPI pattern was derived from the zero-frequency limit of the imaginary part of the Lindhard function, $\chi(q,\omega)$, expressed as:

$$\lim_{\omega \to 0} \frac{\text{Im}[\chi(q,\omega)]}{\omega} \sim \sum_k \delta(\epsilon_k - \epsilon_0) \cdot \delta(\epsilon_{k+q} - \epsilon_0).$$

Here, $\delta(x)$ represents the Dirac delta function, $\epsilon_0$ is the Fermi energy, and $\epsilon_k$ denotes the band energy as a function of wavevector $k$. The vector $q$ corresponds to the Fermi surface scattering vector.

**Extraction of CDW phase maps across domains:**

We employed a two-dimensional lock-in method to extract the information on the CDW phase spatial distributions[62-64]. A topography with CDW modulations can be converted into a modulation image as:

$$M(\boldsymbol{r}) = A(\boldsymbol{r})\cos[\boldsymbol{Q}_i \cdot \boldsymbol{r} + \boldsymbol{\phi}(\boldsymbol{r})]$$

The 2D reference modulation (its modulation frequency $\boldsymbol{Q}_{i=1,2\ or\ 3}$ as the wavevector in reciprocal space) is obtained by the fitting of the peaks in FFT. This procedure gives two orthogonal reference signal which will be used to calculate the relative phase:

$$e^{i\boldsymbol{Q}_i \cdot \boldsymbol{r}} = \cos(\boldsymbol{Q}_i \cdot \boldsymbol{r}) + i \cdot \sin(\boldsymbol{Q}_i \cdot \boldsymbol{r}).$$

The original data is multiplied by the reference signal. A Fourier transform operation is performed and the DC component is obtained with a low-pass filter applied. Then an inverse FFT is performed to obtain the filtered data with only desired frequency.

The vector-based lock-in calculation is done in reciprocal space:

$$M_{\boldsymbol{Q}_i}(\boldsymbol{r}) = \mathcal{F}^{-1}\left\{\mathcal{F}[M(\boldsymbol{r}) \cdot e^{i\boldsymbol{Q}_i \cdot \boldsymbol{r}}] \cdot e^{-q^2/\sigma^2}\right\}$$

Where $\mathcal{F}$ and $\mathcal{F}^{-1}$ are Fourier transform and inverse Fourier transform operations, respectively. $\sigma$ defines the Gaussian cutoff low-pass filter in the momentum space which has the meaning of $1/R$ where R should be carefully chosen to be large enough to contain several modulation periods and not too large than the CDW domain size. The relative propagation phase information can be calculated here:

$$\phi_{\boldsymbol{Q}_i}(\boldsymbol{r}) = tan^{-1}\frac{\mathrm{Im}M_{\boldsymbol{Q}_i}(\boldsymbol{r})}{\mathrm{Re}M_{\boldsymbol{Q}_i}(\boldsymbol{r})}.$$

Due to finite size of the topography and the presence of domains and disorders, there is some uncertainty in locating the exact position of wavevectors for reference signals. This will result in local drift within domains as shown in the phase maps (for example, in Fig. 4d). However, the phase shift across the domain boundary clearly shows an abrupt phase change about $\pm 2\pi/3$, consistently for (1/3, 1/3) in-plane charge order.

**Data availability**

The data that support the findings of this study are available from the corresponding author upon reasonable request.

**Acknowledgements**

We thank Joseph Paddison, Ziqiang Wang, Lingyuan Kong, and Mark Dean for stimulating discussions. This research was supported by the U.S. Department of Energy, Office of Science, Basic Energy Sciences (DOE-BES), Materials Sciences and Engineering Division (ARPES, X-ray and sample growth). ARPES and X-ray scattering measurements used resources at 21-ID-1, 4-ID and 10-ID beamlines of the National Synchrotron Light Source II, a US Department of Energy Office of Science User Facility operated for the DOE Office of Science by Brookhaven National Laboratory (BNL) under contract no. DE-SC0012704. W.R.M. and D.M. acknowledge support from the Gordon and Betty Moore Foundation's EPiQS Initiative, Grant GBMF9069 to D.M. J.I. is supported by NSF Career Award No. DMR-2340394.

STM measurements were supported by the National Science Foundation under grant DMR-2004691. (X.H.) and by the Scan Probe Microscopy FWP under DOE-BES at BNL (A.N.P).

**Author contributions**
F.Y., X.H., and H.T. contributed equally to this work. H.M. conceived and designed the research. F.Y., A.K., A.R., E.V., and H.M. performed the ARPES measurements. F.Y., C.N., and H.M. carried out the X-ray scattering measurements. F.Y., S.K., Y.C., and H.M. performed IXS measurements. X.H., M.T., J.I., and A.P. performed the STM measurements. H.T. and B.Y. performed the DFT calculation and analysis. B.O., W.M., and D.M. synthesized the high-quality single crystals. F.Y., X.H., H.T., and H.M. prepared the manuscript with inputs from all authors.

**Competing interests**
The authors declare no competing interests.

**Additional information**
**Supplementary information** is available at the online version of the paper.
Correspondence and requests for materials should be addressed to Hu Miao, Abhay N. Pasupathy, and Binghai Yan.

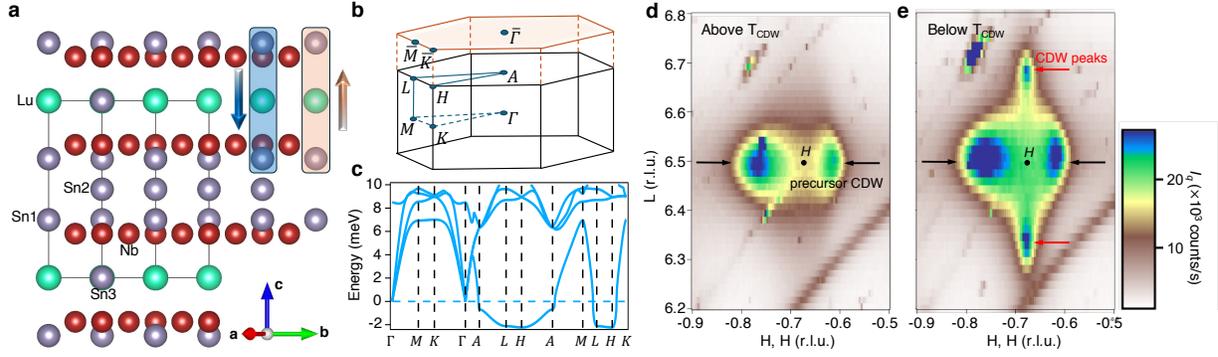

**Figure 1: CDWs in flat phononic kaogme metal LuNb$_6$Sn$_6$. a**, the crystal structure and the out-of-plane distortion of CDW in LuNb$_6$Sn$_6$. **b**, The schematic of BZ. **c**, the calculated phonon dispersion of LuNb$_6$Sn$_6$, which shows a leading-instability at $H$ point with ($k_z=\pi$). **d** and **e**, the intensity plot of diffuse scattering in (H, H, L) plane measured above $T_{CDW}$ (**d**) and below $T_{CDW}$ (**e**). Asymmetric precursor CDW peaks (indicated by black arrows) that are separated away from the H-point are observed T>$T_{CDW}$. The $\sqrt{3}\times\sqrt{3}\times 3$ CDW superlattice peaks (indicated by red arrows) emerge at the $T_{CDW}$.

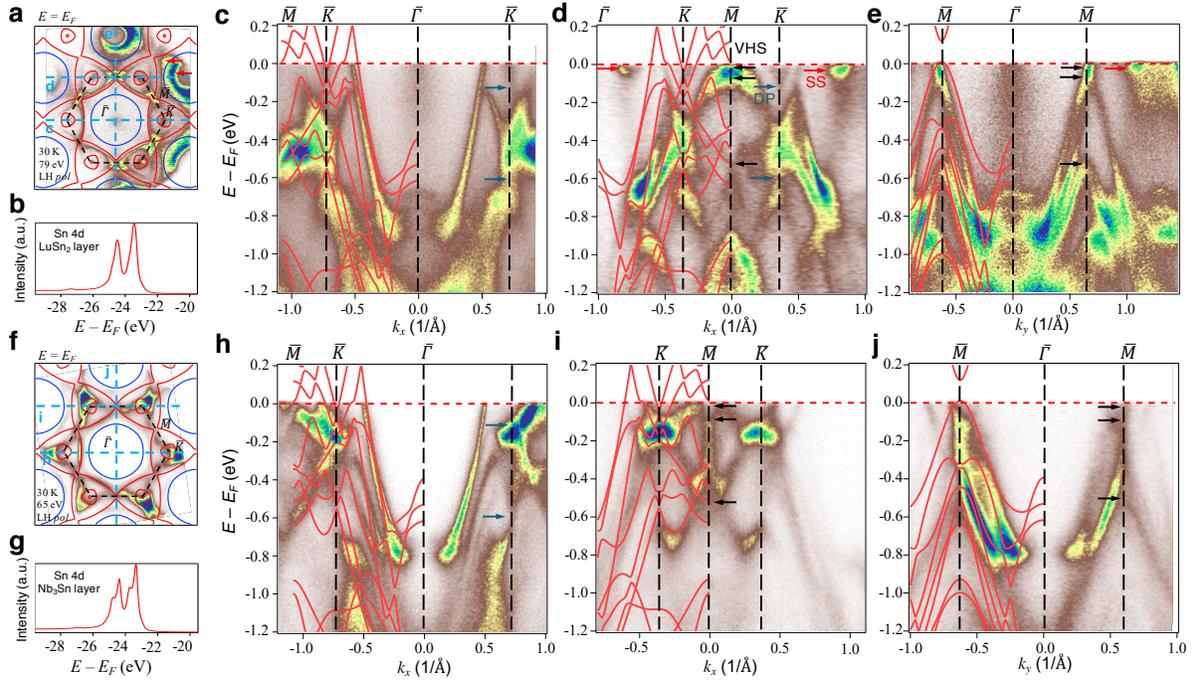

**Figure 2: Electronic structure of LuNb$_6$Sn$_6$. a**, The ARPES determined FS topology on the LuSn$_2$-terminated surface show a triangle-shape FS and small circle pocket around the $\bar{K}$ points, matching well with the DFT calculations. The red and blue lines are the DFT calculated bulk FS at $k_z$=0 and $\pi$, respectively. **b**, The Sn 4d core level measured on the LuSn$_2$-termination shows two peaks corresponding to Sn $4d_{3/2}$ and $4d_{5/2}$. **c-e**, ARPES intensity plots along $\bar{K}$-$\bar{\Gamma}$-$\bar{K}$, $\bar{K}$-$\bar{M}$-$\bar{K}$, and $\bar{M}$-$\bar{\Gamma}$-$\bar{M}$ high symmetry path indicated by the cyan dotted lines in **a**. Red curves are the DFT calculated bulk band at $k_z$=0. The red, blue, and black arrows indicate the surface states (SS), the bulk Dirac points (DP) and the van Hove singularities (VHS), respectively. The data are measured at 30 K and 79 eV photon energy. **f-j**, Same as **a-e**, but measured at the Nb$_3$Sn Kagome termination surface with 65 eV photon energy. The Sn 4d core levels are split into four peaks.

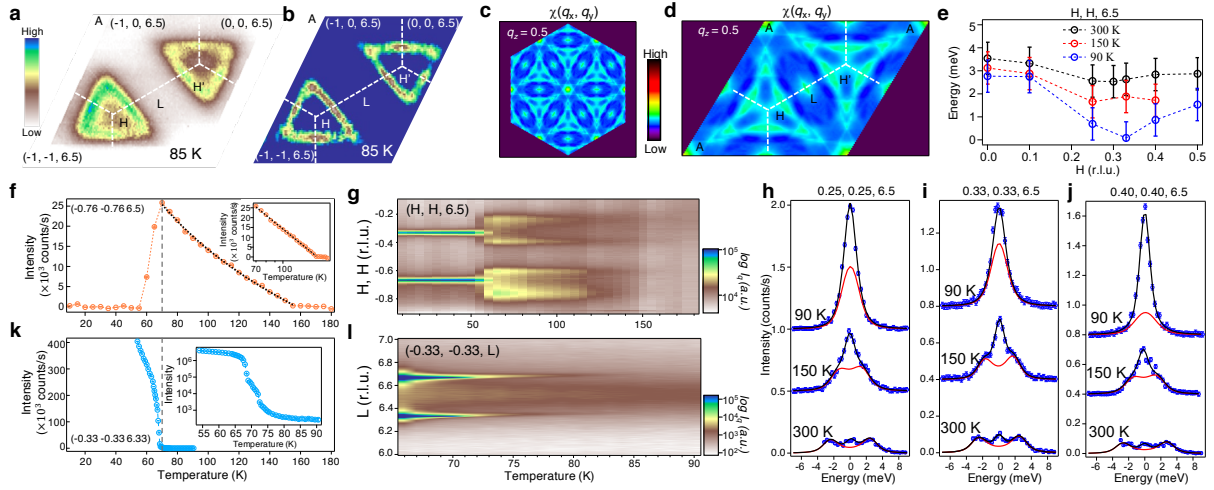

**Figure 3: Temperature dependent evolution of CDWs. a**, the elastic x-ray diffuse scattering in the (H, H, 6.5) plane at 85 K showing 'yield sign'-like hollow triangular pattern. The local intensity maximum is observed at the three corners and centers of the three edges. This pattern is more obvious in **b** that shows the second derivative plot of **a**. The white dotted lines label the 2D BZ in the $L$= 6.5 plane. **c**, the calculated charge susceptibility, $\chi(q)$, at $q_z = 0.5$ plane. **d**, zoomed in plot of $\chi(q)$ displays the same 'yield sign' pattern shown in **a** and **b** with stronger intensity at edge centers and corners. **e**, IXS determined phonon dispersion at 300, 150, and 90 K, revealing a complete phonon softening at $Q$=(0.33, 0.33, 6.5). The error bars for the phonon dispersions represent the energy resolution of IXS. The fitting error bars are much smaller. **f**, the temperature dependent diffuse scattering intensity at (-0.76, -0.76, 6.5). The intensity monotonically increases between 160 K and $T_{CDW}$. Below the $T_{CDW}$, the intensity is quickly suppressed and reaches background with 10 K temperature window. The cyan line shows the logarithmic fitting of the data. The inset shows the same data but plotted as $ln$T scale for the horizontal axis. The gray dotted line marks $T_{CDW}$. **g**, The intensity plot of temperature dependent diffuse scattering along (H, H, 6.5). Note the intensity is shown in the log scale. **h-i**, the IXS raw data at $T$=300 , 150, and 90 K, at $Q$=(0.25, 0.25, 6.5), (0.33, 0.33, 6.5) and (0.40, 0.40, 6.5), respectively. **k**, the temperature dependent CDW superlattice intensity at (-0.33, -0.33, 6.33). The inset shows the intensity in the log-scale. **l**, The colormap of CDW superlattice scan along (-0.33, -0.33, L) vs temperature. Note the intensity is shown in the log scale.

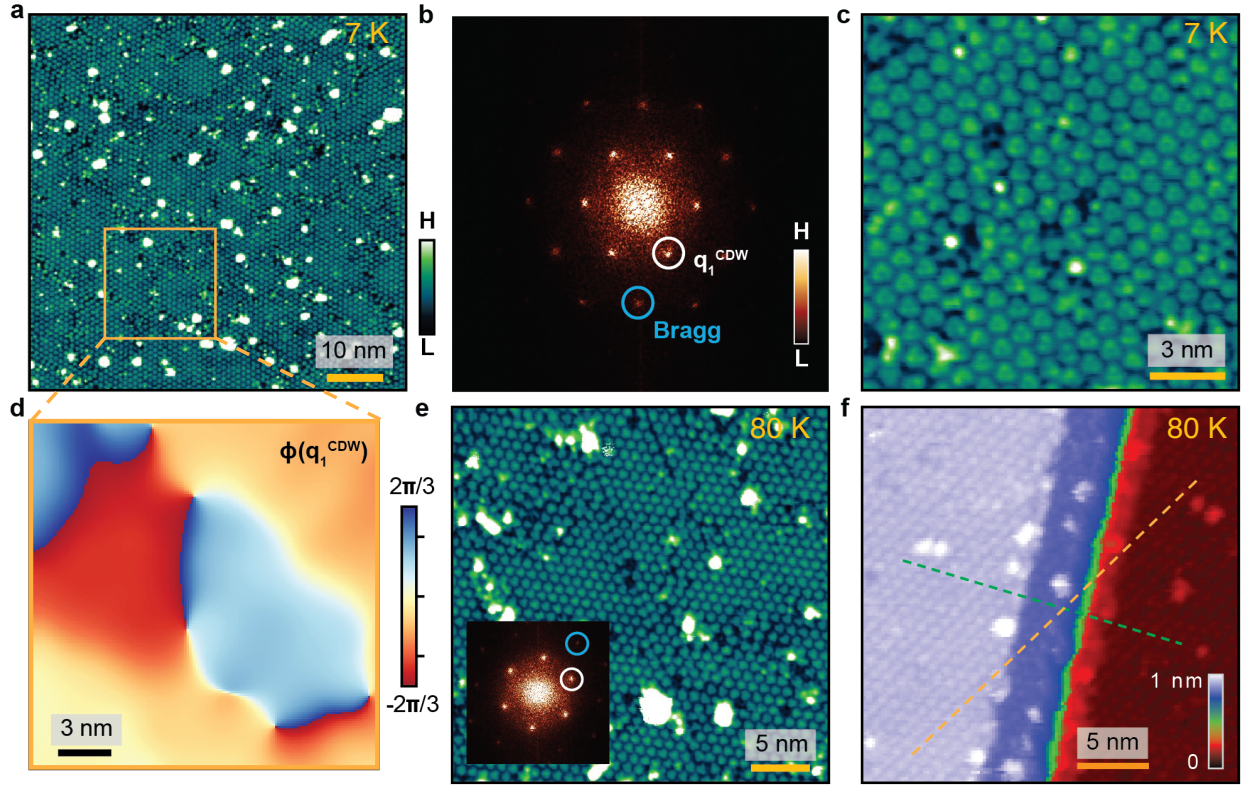

**Figure 4: Real space imaging of short-ranged CDWs**. **a**, STM topography measured on Kagome Nb$_3$Sn-terminated surface at low-temperature (7 K) showing the charge order deep in the CDW phase. **b**, FFT of **a**. The CDW peaks are observed at $\boldsymbol{Q}_{//}=(1/3, 1/3)$ (white circle) in the reciprocal space. **c**, Zoom-in topography of the low-T CDW reveals triangular trimers. **d**, The spatial phase map of $q_1$ (marked in **b**) for the highlighted box region in **a**. The phase difference for domains is about $\pm 2\pi/3$. Spatial phase maps for the other two wavevectors can be found in Supplementary Materials. **e**, STM topography measured at 80 K>$T_{CDW}$ reveals precursor CDW. Insets shows FFT of **b**. **f**, STM topography of a precursor CDW domain over one unit-cell step edge. Marked by the dashed lines, the precursor CDW exhibits $\pi$ phase shift along the two propagating directions, consistent with x-ray diffraction.